\documentclass[apjl,twocolumn]{openjournal}
\usepackage{natbib} 
\usepackage[dvipsnames]{xcolor}
\usepackage{aas_macros} 
\usepackage{amssymb}
\usepackage{amsmath}
\usepackage{threeparttable}
\usepackage[dvipsnames]{xcolor}
\usepackage[title]{appendix}
\usepackage[hidelinks]{hyperref}	
\hypersetup{colorlinks=false,linkcolor=false,citecolor=false,filecolor=false,urlcolor=false}
\usepackage[caption=false]{subfig}
\usepackage{soul}                          

\newcommand{\revision}{}
\newcommand{\revis}{}

\newcommand{\kms}{km s$^{-1}$} 

\newcommand{\gcsub}{GC$_{\rm sub}$}
\newcommand{\gcnon}{GC$_{\rm non}$}

\begin{document}
\title[]{Rotational Kinematics in the Globular Cluster System of M31: \\ Insights from Bayesian Inference\vspace{-15mm}}
\author{Yuan (Cher) Li$^{1}$\thanks{$^*$E-mail: \href{mailto:yli464@aucklanduni.ac.nz}{yli464@aucklanduni.ac.nz}}, Brendon J. Brewer$^{1}$, Geraint F. Lewis$^{2}$,
Dougal Mackey$^3$
}

\affiliation{$^{1}$Department of Statistics, The University of Auckland, Private Bag 92019, Auckland 1142, New Zealand}
\affiliation{$^2$Sydney Institute for Astronomy, School of Physics, A28, The University of Sydney, NSW 2006, Australia}
\affiliation{$^3$Independent researcher, Charnwood, Canberra, ACT 2615, Australia}

\begin{abstract}
As ancient stellar systems, globular clusters (GCs) offer valuable insights into the dynamical histories of large galaxies. Previous studies of GC populations in the inner and outer regions of the Andromeda Galaxy (M31) have revealed intriguing subpopulations with distinct kinematic properties. Here, we build upon earlier studies by employing Bayesian modelling to investigate the kinematics of the combined inner and outer GC populations of M31. Given the heterogeneous nature of the data, we {\revision examine} subpopulations defined by {\revision GCs'} metallicity and {\revision by} associations with substructure, in order to characterise possible relationships between the inner and outer {\revision GC} populations. We find that lower-metallicity GCs and those linked to substructures exhibit a common, more rapid rotation, whose alignment is distinct from that of higher-metallicity and non-substructure GCs. Furthermore, the higher-metallicity GCs rotate in alignment with Andromeda’s stellar disk. These pronounced kinematic differences reinforce the idea that different subgroups of {\revision GCs} were accreted to M31 at distinct epochs, shedding light on the complex assembly history of the galaxy.
\end{abstract}

\section{Introduction}
\label{Intro}
Formed in the early epochs of the universe, globular clusters (GCs) offer valuable insights into the properties and formation of galaxies \citep{2018RSPSA.47470616F,2020rfma.book..245B}. Given its proximity, the GCs population of the Andromeda Galaxy (M31) has the promise of revealing the accretion history of a large galaxy, shedding light on galactic growth in the local environment, including the life of our own Milky Way. 

Separate studies of the kinematics and spatial distribution of {\revis GCs} in the outer and inner regions of Andromeda have {\revis revealed a complex picture of distinct accretion events, some relatively recent and others more ancient}  \citep[e.g.][]{2019Natur.574...69M,2023MNRAS.518.5778L}. Intriguingly, these suggest a possible causal relationship between the structures found in the outer and inner halo. {\revision We therefore aim to} investigate the kinematic characteristics of M31's combined GC population (inner and outer GCs)
to {\revision assess} whether this {\revision combined approach} supports a {\revision coherent picture of Andromeda's accretion history.} 


At $\sim765$ kpc, M31 is the closest large neighbour to the Milky Way \citep{riess2012cepheid}, and with a mass of $1.5 \times 10^{12} {\mathrm M_{\sun}}$ \citep{2014MNRAS.443.2204P}, M31 is $\sim30 \%$ more massive.
{\revis Almost 500 GCs associated with M31 have been identified out to $\sim150$ kpc \citep[e.g.][]{2001AJ....122.2458B,2010ApJ...717L..11M}. This substantial GC system likely reflects not only M31’s higher mass but also its more complex assembly and accretion history, which distinguishes it from the Milky Way. This is significantly
larger than the Milky Way's GC population, consistent with the larger mass
of M31 and the substantial intrinsic scatter in the relation between
GC population size and host galaxy mass \citep{harris2013catalog}.}
This rich population of GCs has long attracted considerable interest \citep{1985ApJ...288..494C,1988ApJ...333..594E}, with \citet{2007ApJ...655L..85M} comparing the features of M31's and the Milky Way's outer halo (15 kpc $\lesssim$  $R_p$ (projected radii) $\lesssim$ 100 kpc) GCs. Their findings imply that the outer halo of M31 has a greater number of metal-poor, {\revision more centrally concentrated}, and highly luminous GCs than the Milky Way. Many other studies have compared the properties between the Milky Way and M31 \citep[e.g.][]{1997MNRAS.285L..31K,2007ApJ...671.1591I,2004AJ....128.1623B,2004ApJ...614..158B,2012AJ....143...14S}. These point to M31's accretion history differing from the Milky Way's,
potentially involving complex merger histories \citep{akib2025impactmergerhistoriestiming,akib2025intriguing}. 

\citet{2010ApJ...717L..11M} modelled the spatial distribution of GCs in the halo of M31 using a sample of 61 observed GCs from PAndAS \citep[the Pan-Andromeda Archaeological Survey;][]{mcconnachie2009remnants}. Focusing primarily on GCs beyond $R_p = 30$ kpc (M31’s outer halo), they found that these clusters are associated with underlying tidal debris features and exhibit a roughly isotropic spatial distribution. This finding was later confirmed by \citet{2012ASPC..458..275M}, who demonstrated that the correlation between outer halo GCs and tidal debris persists when considering the full PAndAS sample. Furthermore, \citet{2012ASPC..458..275M} asserted that over 80\% of M31’s outer halo GCs were acquired through the accretion of satellite host galaxies. However, the findings of \citet{2019MNRAS.484.1756M} suggest that, at $R_p > 25$ kpc, approximately 35\% to 62\% of GCs show signs of having been accreted onto M31’s halo. This discrepancy may arise from \citet{2012ASPC..458..275M} analysing only {\revision 79 GCs} of M31’s {\revision outer halo compared to the 92 outer halo GCs studied by \citet{2019MNRAS.484.1756M}}.

\begin{table*}
	\centering
	\caption{Prior probability distributions for the unknown parameters for Model 1.}
	\label{prior}
	\begin{tabular}{lll} 
		\hline
		Parameter & Description & Prior\\
		\hline
            $\sigma$ (\kms)& Velocity dispersion & Uniform(0, 1000)\\
		      $A$ (\kms) & Rotational 
            Amplitude & Uniform(0, 1000) \\
            $\phi$ (degrees) & Orientation of rotation axis & Uniform($-180$, $180$) 
	\end{tabular}
\end{table*}

\begin{figure}
\centering
\includegraphics[width=0.9\linewidth]{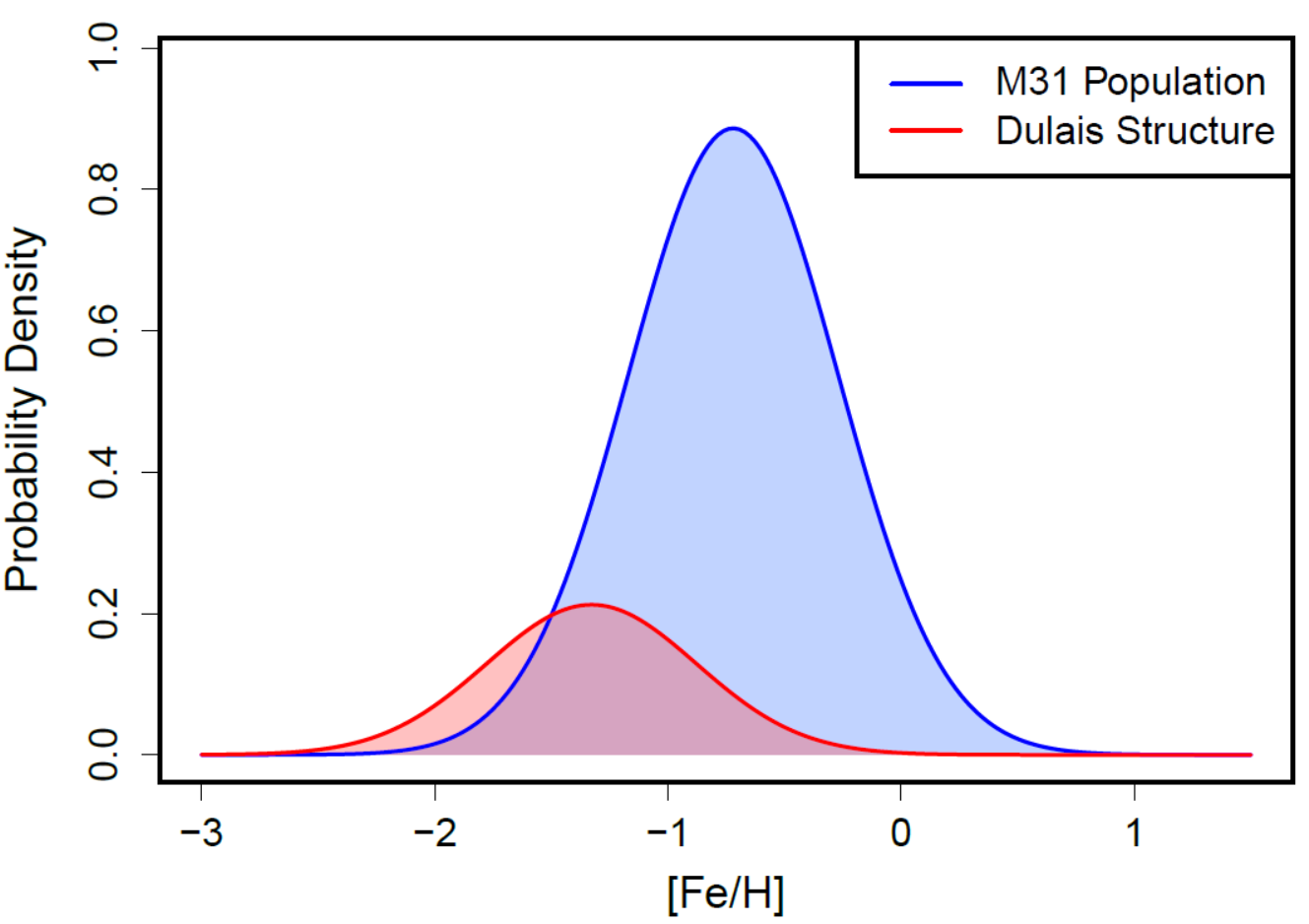}
\caption{{\revis Schematic representation of the metallicity distribution of M31’s inner globular cluster population ($R_p < 25$ kpc, shown in blue) and of the Dulais Structure subset (shown in red), identified by \citet{2023MNRAS.518.5778L} using the lowest metallicity inner GCs. While the populations would actually overlap
as illustrated here, our modelling uses the simplifying assumption that
the overall GC population is neatly split into two at a critical value
of metallicity.}}
\label{mixture}
\end{figure}

\begin{figure*}
\centering
\includegraphics[width=0.9\linewidth]{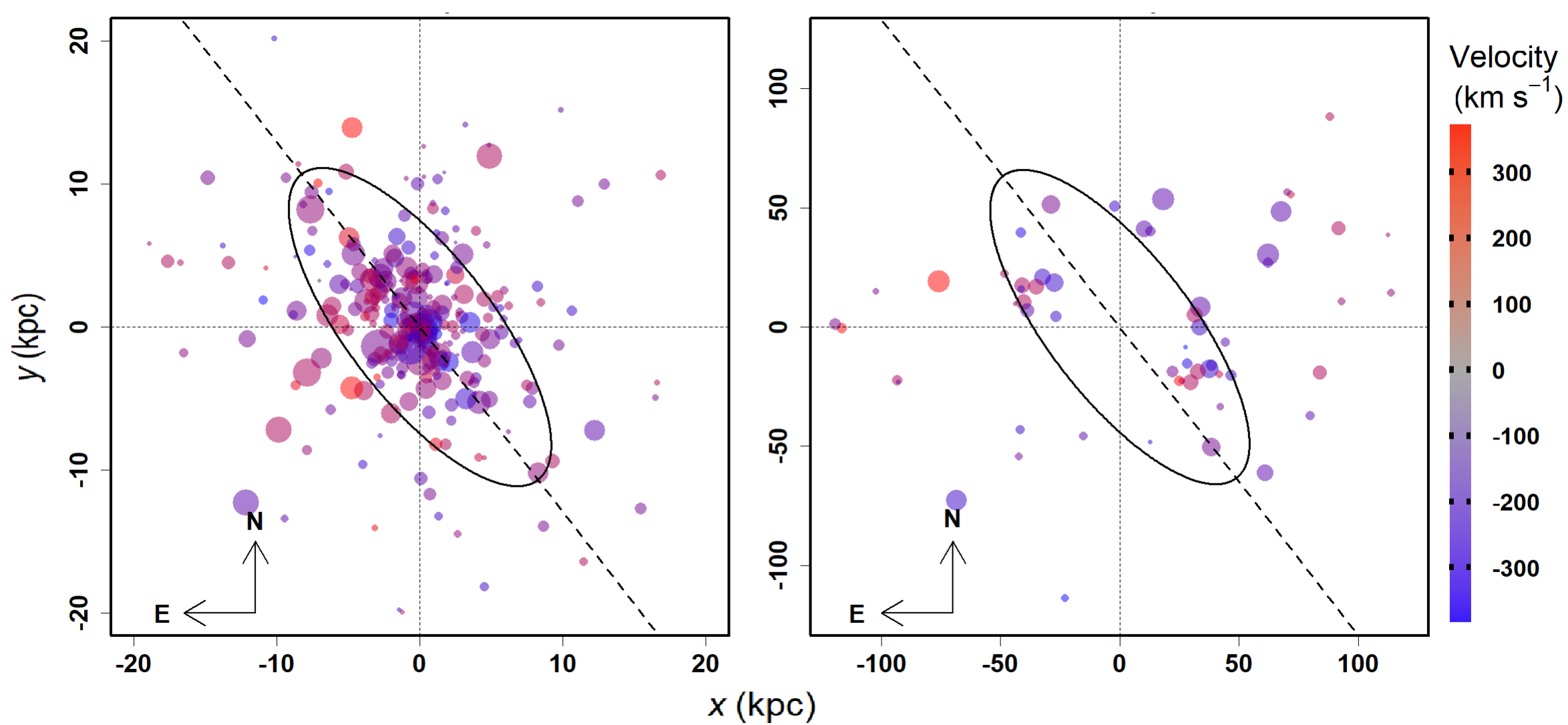}
\caption{{\revision Spatial distribution of M31 GCs, {\revis studied in this paper, separated into inner ($R_p < 25$ kpc; left panel) and outer ($R_p \ge 25$ kpc; right panel) populations.} Points are colored according to line-of-sight velocity relative to the galaxy, with red indicating positive velocities and blue indicating negative velocities. The size of each point is scaled by the absolute value of the velocity. Black ellipses indicate the disk orientation of the M31 galaxy. Compass arrows indicate the north (N) and east (E) directions.}}
\label{combined}
\end{figure*}

\citet{2014MNRAS.442.2929V} demonstrated that the outer halo GC population (those at $R_p > 30$ kpc) rotates along the same axis as the stellar disk of M31, albeit with a lower amplitude of $86 \pm 17$ \kms. {\revision For comparison, the disk rotation velocity is $\sim 226$\,\kms~between 20 and 35 kpc \citep{2006ApJ...641L.109C}, indicating that the outer halo GCs rotate at roughly 38\% of the disk's rotational speed.} Additionally, they identified a decline in velocity dispersion as a function of projected distance from M31’s centre, providing further evidence of the complex kinematic structure of the galaxy’s halo.
\citet{2019Natur.574...69M} investigated how the rotational properties of outer halo (GCs) in M31 vary depending on their association with substructures. Their findings revealed that GCs linked to substructures and those unassociated with them rotate in perpendicular orientations. There is compelling evidence \citep{2019MNRAS.484.1756M,2019Natur.574...69M} suggesting that GCs not located on substructures share similarities with ancient dwarf galaxies that accreted into M31’s halo approximately 12 {\revision Gyr} ago \citep{2008ApJ...689..936J}. In contrast, GCs associated with substructures are likely to have been accreted more recently \citep{2018ApJ...868...55M}, {\revis Gyr} after the initial accretion events of the non-substructure GCs \citep{2019Natur.574...69M}.

\begin{figure}
\centering 
\includegraphics[width=8.8cm]{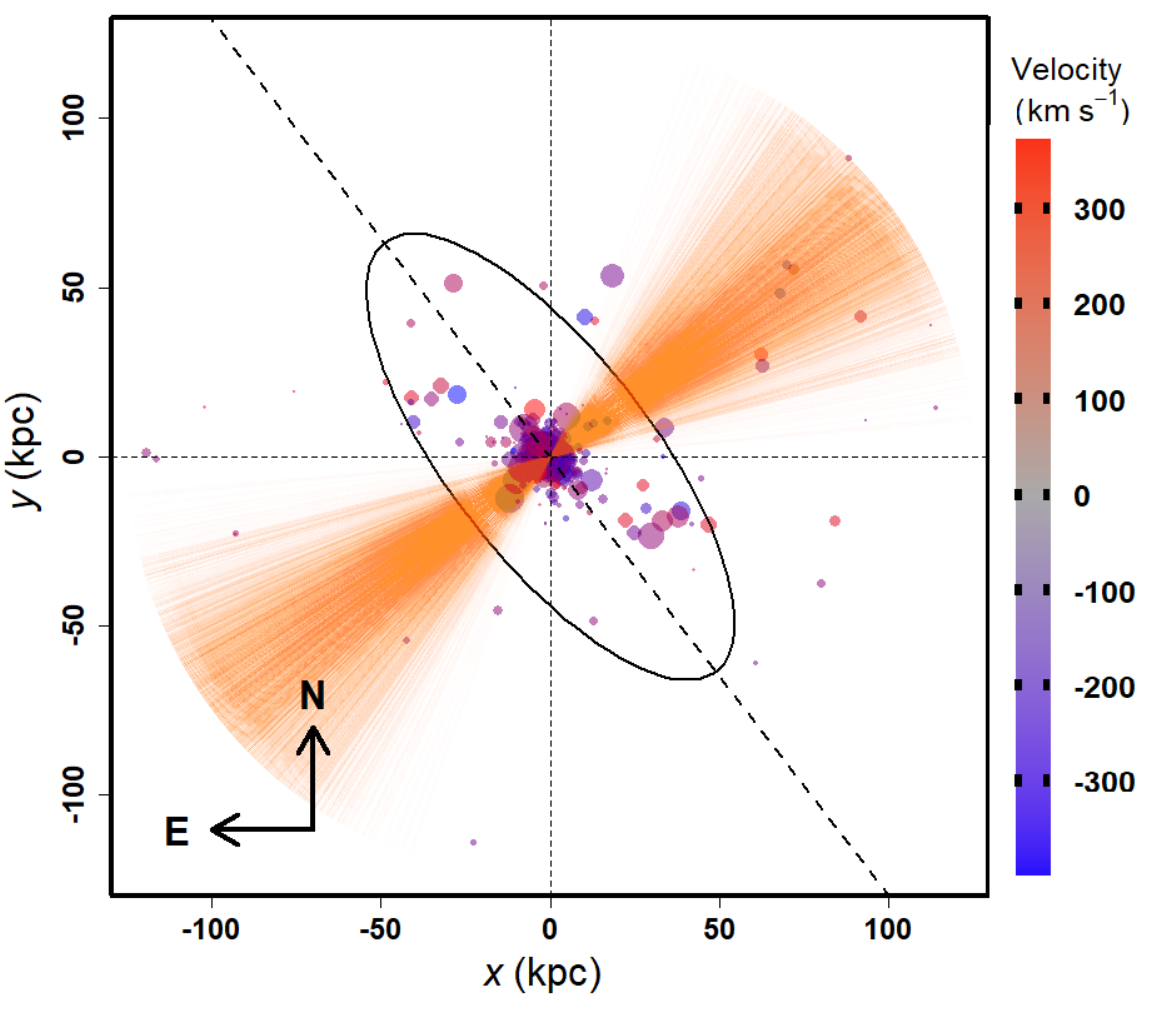}	
\caption{The M31 GC population, colour-coded
by radial velocity in the M31 frame. The absolute value of the line-of-sight velocity is also indicated
by the size of each point. The orange lines are representative samples of the orientation angle drawn from the parameter exploration of Model 1. The ellipse represents M31's disk orientation.}
\label{vm1}
\end{figure}

Beyond studies of the outer halo GCs, investigations of inner halo GCs have also provided critical insights into M31’s structure. \citet{2016ApJ...824...42C} conducted a comprehensive spatial-kinematic-metallicity analysis of GCs within a restricted radius of $R_p = 21$ kpc. By classifying GCs into three groups based on metallicity, they concluded that {\revision the system of} higher metallicity GCs ([Fe/H] $> -0.4$) exhibit rotational {\revis velocity}, velocity dispersion, and spatial properties akin to M31’s stellar disk. Conversely, {\revision the system of} lower metallicity GCs ([Fe/H] $< -1.5$) were the least correlated with the disk, displaying moderately prograde rotation and a higher velocity dispersion. Furthermore, \citet{2023MNRAS.518.5778L} found that certain rotational patterns of inner halo GCs (those with a lower metallicity --- part of what they called the `Dulais structure' after a Welsh river) align closely with those of outer halo GCs, suggesting a {\revision common progenitor or origin. The metallicity distribution of the inner GCs and those associated with the Dulais Structure is illustrated schematically in Figure~\ref{mixture}, showing that the latter dominates at low metallicity}. {\revis While the two populations would overlap in reality,
our modelling assumptions simplify matters by assuming that a critical
metallicity value neatly divides the GC population into two parts,
following \citet{2023MNRAS.518.5778L}.}

\begin{table}
	\centering
	\caption{Summary table of parameter estimates for Model 1.} 
	\label{vm1sum}
	\begin{tabular}{lr} 
		\hline
		Parameter & Estimate\\
		\hline 
	  $\sigma$ (\kms) & $137.66^{+5.54}_{-5.11}$ \\[0.5em]
		$A$ (\kms)& $68.74^{+9.95}_{-10.31}$ \\[0.5em]
        $\phi$ (degrees)& $37.45^{+9.73}_{-9.58}$\\[0.5em]
        {\revision PA (degrees)} & {\revision 307.45 $^{+9.73}_{-9.58}$}\\[0.5em]
        $A/\sigma$ ratio & $0.50^{+0.07}_{-0.08}$ \\[0.5em]
		\hline
	\end{tabular}
\end{table}

The structure of the paper is as follows: Section~\ref{Data} describes the data used in this current study. Section~\ref{MB} discusses the kinematic models applied to examine the GC population of the M31 galaxy, with Bayesian inference used to determine the posterior probability distributions for the model parameters and the models’ marginal likelihoods (evidence). Finally, the discussion and conclusions are presented in Section~\ref{Dis}.

\section{Data}
\label{Data}
The dataset used in this paper, which initially contained 422 GCs, is from a {\revision publicly available} catalogue \citep{2016ApJ...824...42C}. Of the 422 clusters, 345 are from M31's inner region ($R_p < 25$ kpc), whereas 77 are from the outer region. 
The data encompassing {\revision a majority of the} known inner GCs of M31 came from spectroscopic studies with the Hectospec multifibre spectrograph \citep{2005PASP..117.1411F} on the 6.5-meter MMT Observatory telescope in Arizona. Using these spectra, \citet{2016ApJ...824...42C} produced metallicity estimates and a collection of uniformly determined line-of-sight velocities for roughly 94\% of GCs with projected radii within 21 kpc. During data collection, {\revision the metallicities of six inner globular clusters were not measured and were therefore flagged as $-99$}; as a result, we excluded these six {\revision GCs}. We also decided to remove those GCs with metallicity $\geq$ \text{-}0.4 as these are most likely associated with the disk of M31 \citep{2016ApJ...824...42C}, and also so we would
have a dataset consistent with previous research \citep{2023MNRAS.518.5778L}. After removing these data points, the metallicity of the remaining inner GCs ranges from $-2.8<[\textnormal{Fe}/\textnormal{H}]<-0.4$. The inner GC data were obtained through MMT spectroscopy and were originally published by \citet{2016ApJ...824...42C}.

The 77 outer GCs, which lack metallicity measurements, have been categorized as either not associated with a substructure (a subgroup of GCs henceforth referred to as \gcnon), associated with a substructure (henceforth referred to as \gcsub), or ambiguous.  In \citet{2019Natur.574...69M}, the 19 ambiguous clusters were included in the analysis,
and an attempt was made to infer their substructure status from the data,
based on whether their other properties are more consistent with the \gcsub~or
\gcnon~subpopulations. 
As our analysis is more complex in other ways {\revision (see Section \ref{MB})},
we excluded these ambiguous clusters to limit the complexity of our approach. {\revision Hence, in total, 26 \gcnon~and 32 \gcsub~outer globular clusters were analysed in this study.}
Note that, due to the complexities of stellar structure in the inner regions of M31, it is not possible to unambiguously associate the inner GCs with distinct stellar substructure. 
It should be noted that photometric metallicity estimates are now available (and a few spectroscopic ones) for the outer GCs \citep{2025MNRAS.542L..60M}; however, we have opted to rely on the complete spectroscopically derived metallicity estimates available for the inner population, {\revision as the outer GC metallicities are heterogeneous in origin and not uniformly calibrated.} The outer GC data were obtained from PAndAS \citep{2018ApJ...868...55M} and can be downloaded from the Canadian Astronomical Data Center (\href{http://www.cadc-ccda. hia-iha.nrc-cnrc.gc.ca/en/community/pandas/query.html}{CADC}).

{\revis In summary,} after combining the datasets from the inner (339) and outer (77) regions, we removed 19 ambiguous outer clusters, six inner clusters without metallicity measurements, and all clusters with metallicity [Fe/H] $\geq -0.4$. The final dataset used therefore comprises 336 GCs.
{\revis The spatial distribution of the inner and outer GCs
that we used in this study
can be seen in Figure \ref{combined}.}

\begin{figure}
\centering
\includegraphics[width=\linewidth, height = 8cm]{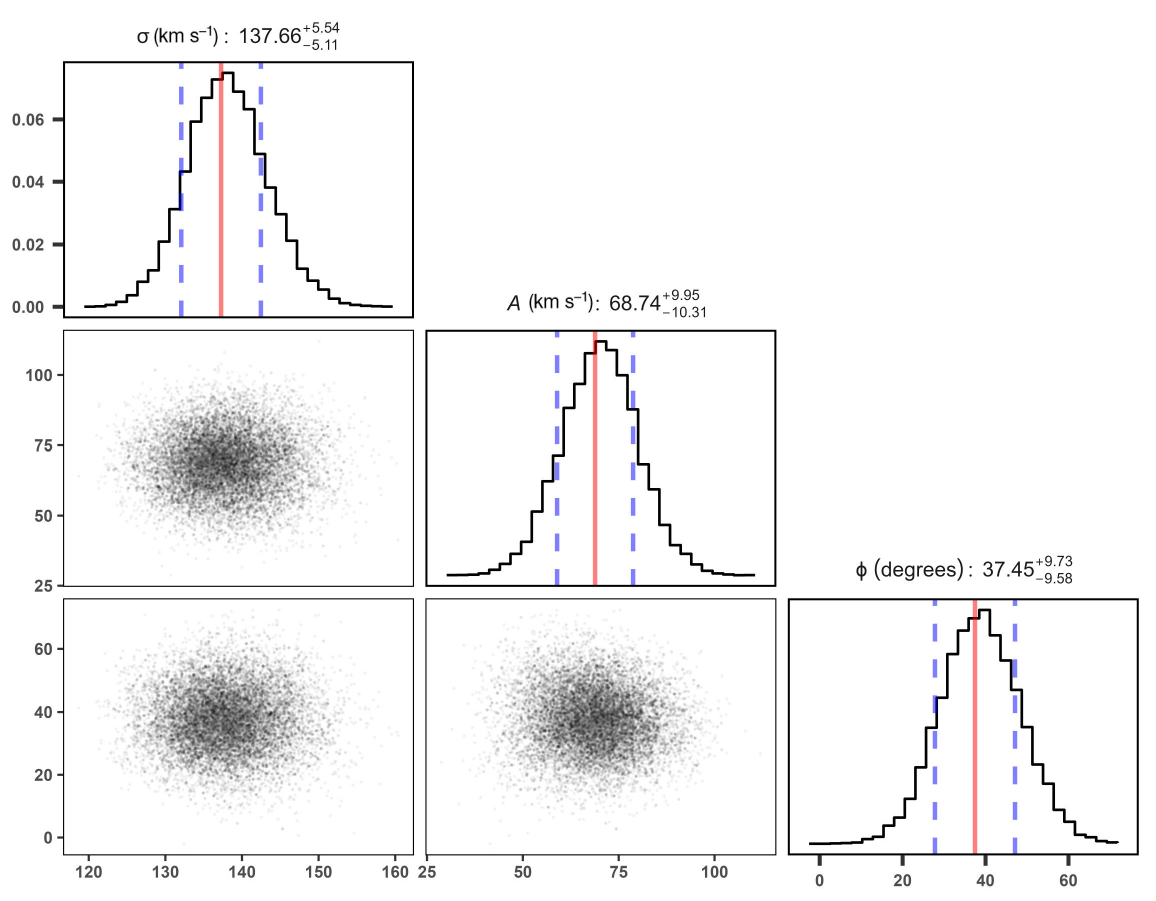}
\caption{Corner plot \citep{corner} of the posterior distribution for Model 1's parameters.
There are no strong correlations or other dependencies in the posterior
distribution, and all of the marginal distributions are normal to
a good approximation. {\revision The red solid line represents the median (50th percentile), whereas the blue dashed lines indicate the 16th and 84th percentiles of the posterior distribution of the parameters.}} 
\label{vm1hist}
\end{figure}

\section{Kinematic Models and Bayesian Inference}
\label{MB}
Upon combining the inner and outer GC data, we first investigated the overall rotational orientation of M31's GCs using a simple kinematic model,
which we call Model 1. This model {\revision is comprised of only a singular component describing all GCs which}~serves to capture the
overall rotational pattern of the GCs, and to provide a simple baseline
for model comparison purposes.
The kinematic model employed gives
the line-of-sight velocity $v$ as a function of position on the sky
$(x,y)$, using the following functional form:
\begin{equation}\label{eqModel1}
v(x, y) = A \sin(\theta - \phi),
\end{equation}
where $(x, y)$ {\revision are the projected coordinates of the GC on the sky, measured relative to the centre of M31},
the unknown parameter $\phi$ indicates the orientation of the GCs' rotation axis, and $A$ is the amplitude of the rotational velocity.
{\revision For the implementation of the analysis, $\phi$ was defined using
the mathematical convention with respect to the $x$ and $y$ axes as displayed
in Figure~\ref{combined}. For convenience, in the results we present
$\phi$ along with the standard position angle (PA) defined from north through
east.}

The quantity $\theta$ is the angular coordinate corresponding to
$(x, y)$ when translated into plane polar coordinates.
We also assume that the single component has a spatially
flat velocity dispersion $\sigma$, also unknown, which appears in the likelihood
function, defined in Section~\ref{P&L}. {\revision Note that the observed line-of-sight velocities are converted to M31-centric velocities.}
The chosen functional form (Equation \ref{eqModel1})
is known as the $V$ model, after \citet{2014MNRAS.442.2929V}. However, alternative functional forms are possible.
In this study, we also utilized two additional model types,
$F$ and $S$, whose details are given in Appendix~\ref{Models}.

\subsection{Prior Distributions and the Likelihood Function}
\label{P&L}
Table \ref{prior} summarizes the prior distributions for the parameters.
In Model 1, there
are only three unknown parameters: amplitude $A$, orientation $\phi$, and
velocity dispersion $\sigma$. {\revision It should be noted that while $A$ and $\phi$ appear explicitly in Equation \ref{eqModel1} as part of the line-of-sight velocity model, $\sigma$ enters the likelihood function (Equation \ref{likelihood}) through the variance term, representing the intrinsic velocity dispersion of the GC system.}

{\revision The velocity dispersion, denoted by $\sigma$, quantifies the random motions of the GCs about the modelled systemic rotation. Physically, it reflects the degree to which individual cluster velocities deviate from the mean rotational trend, providing a measure of the dynamical “temperature” or kinematic spread of the system. In the context of our models, $\sigma$ is included as a parameter in the likelihood function to account for this intrinsic scatter in the observed line-of-sight velocities. A larger $\sigma$ indicates a more dynamically hot or pressure-supported system, whereas a smaller $\sigma$ (relative to $A$) corresponds to a more rotationally supported population.}

For simplicity, we have used uniform priors. However, since we intend
to carry out Bayesian model comparison, the limits of the uniform priors
need to be considered, as the results will be sensitive to these choices.
For the parameters $A$ and $\sigma$, we have imposed an upper bound of 1000 \kms, to rule out astrophysically implausible values. For the orientation
angle $\phi$, the uniform distribution arises 
as a natural prior from considerations of rotational symmetry and is
automatically bounded.

The probability distribution for the measured line-of-sight velocities $v_{i}$ given the parameters, which will also form the likelihood function, is assumed to be a normal distribution with mean $v(x_i, y_i)$ and a variance of $\sigma^2 + s_i^2$, where
$s_i$ is the reported error bar on velocity $i$.
The likelihood function, assumed to be a normal distribution, is given by
\begin{equation}
\mathcal{L}(A, \phi, \sigma) = \prod_{i=1}^n \frac{1}{\sqrt{2\pi(\sigma^2 + s_i^2)}} \exp \left(-\frac{(v_i -v(x_i, y_i))^2}{2(\sigma^2 + s_i^2)}\right).
\label{likelihood}
\end{equation}

\subsection{Bayesian Inference}
\label{BI}
Throughout this paper, we calculate posterior distributions for the parameters
using Bayesian inference. The posterior distribution
for any model parameters $\omega$ is obtained by Bayes's theorem, given by
\begin{equation}\label{M31br}
\mathcal{P}(\omega \,|\, \mathcal{D}) = \frac{\mathcal{P}(\omega)\mathcal{P}(\mathcal{D}\,|\,\omega)}{\mathcal{P}(\mathcal{D})}.
\end{equation}
Here, $\omega$ represents a vector of unknown parameters\footnote{Unknown parameters are traditionally called $\theta$,
but this has already been used for the angular coordinate on the sky.}.
In Equation \ref{M31br}, $\mathcal{D}$ represents the data, $\mathcal{P}(\omega)$ represents the prior probability distribution of the parameters, $\mathcal{P}(\mathcal{D}|\omega)$ represents the likelihood function, and lastly, $\mathcal{P}(\mathcal{D})$ represents the marginal likelihood value, also known as the evidence.

\begin{figure*}
\centering
\includegraphics[width=\linewidth]{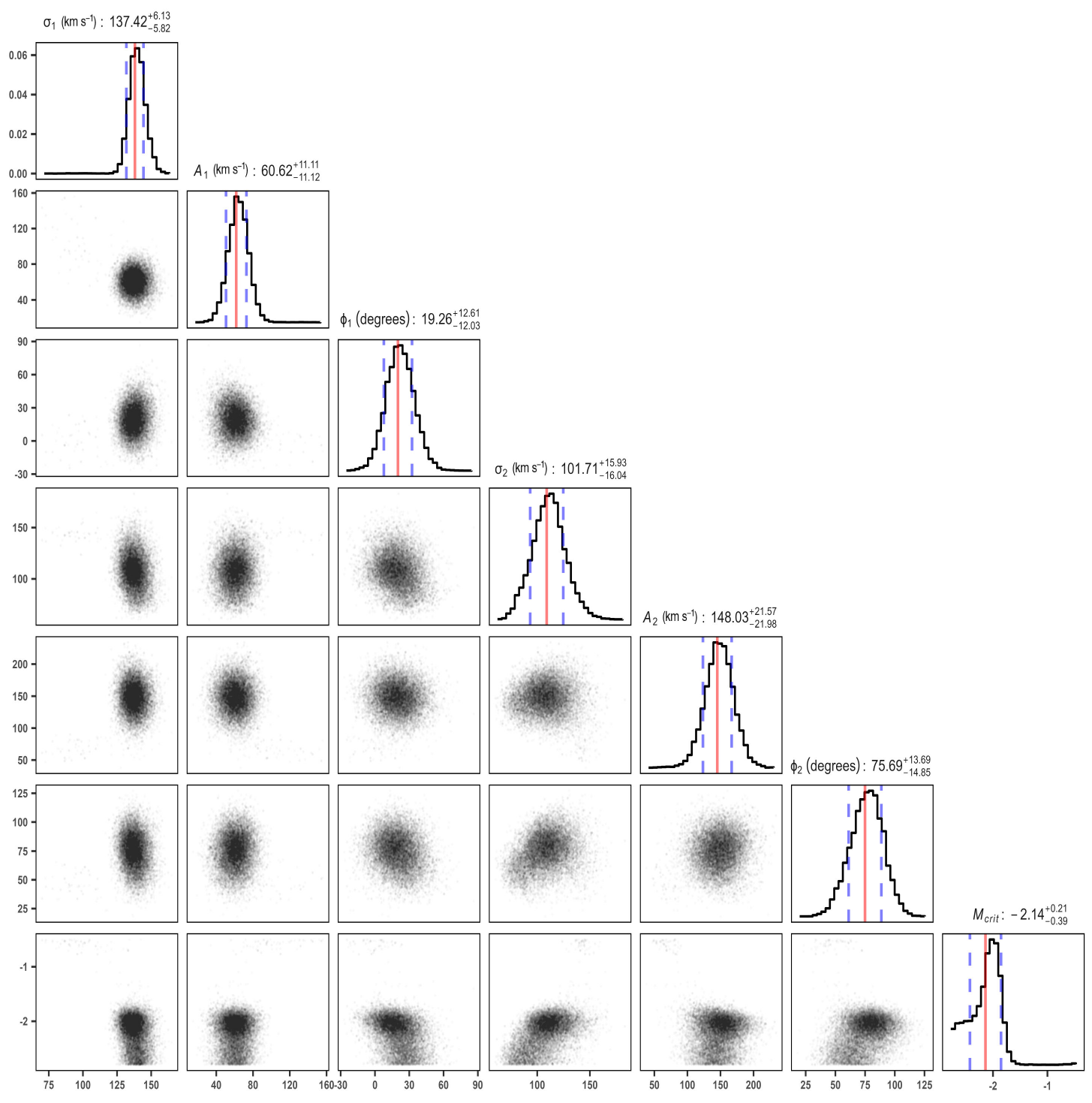}
\caption{The posterior distribution for parameters of Model 2.1. Here, $\sigma_1$, $A_1$ and $\phi_1$ are the components for \gcnon~and higher metallicity GCs, while $\sigma_2$, $A_2$ and $\phi_2$ are the components for \gcsub~and lower metallicity GCs.}
\label{vms2hist}
\end{figure*}

In this paper, Nested Sampling \citep{2004AIPC..735..395S, ashton2022nested}
and Diffusive Nested Sampling \citep{dns, dnest4} were used to generate samples from the posterior distribution and to determine the marginal likelihood or evidence of the model, expressed as follows:
\begin{align}
\mathcal{P}(\mathcal{D}) &= \int \mathcal{P}(\omega)\mathcal{P}(\mathcal{D}\,|\,\omega) \, d\omega
\end{align}
or alternatively
\begin{equation}\label{M31NS}
\mathcal{Z}  = \int \mathcal{L}(\omega) \pi (\omega) \,d\omega,
\end{equation} where the prior distribution is denoted by $\pi(\omega)$, the likelihood function is denoted by $\mathcal{L}(\omega)$, and the evidence or marginal likelihood is denoted by $\mathcal{Z}$. 
Once the marginal likelihoods of the models are determined through nested sampling, we can select the most plausible model based on the highest marginal likelihood value, or we can propagate uncertainty about the model
into any final conclusions using probability theory. 

Finally, the Bayes Factor, the ratio of the marginal likelihoods of two models, can be determined. It is another way of expressing conclusions which could be
expressed as model probabilities.
The Bayes Factor describes the degree to which the data favours one model over the other. The Bayes Factor for comparing some model $\mathcal{M}_1$
to another $\mathcal{M}_2$ is represented by the following equation:
\begin{equation}\label{M31BF}
\mathcal{B}\mathcal{F}(\mathcal{M}_1,\mathcal{M}_2) = \frac{\mathcal{P}(\mathcal{D}\,|\,\mathcal{M}_1)}{\mathcal{P}(\mathcal{D}\,|\,\mathcal{M}_2)},
\end{equation}where $\mathcal{M}_i$ denotes the model.

\subsection{Metallicity Uncertainties}
{\revision Although the models discussed so far do not incorporate metallicity information, other models (described in Section \ref{m&s}) make use of the GCs’ metallicities, and therefore metallicity uncertainties must be considered. As mentioned in the Section~\ref{Data}, six inner GCs without metallicity measurements and those with metallicity [Fe/H] $\geq -0.4$ have been removed; the final sample of inner GCs with measured metallicities comprises 278 objects. Since the uncertainties}
are provided for the metallicity measurements of the 278 inner GCs, we adopt the same strategy as \citet{2023MNRAS.518.5778L}. This introduces 278 additional parameters {\revision  to the model}, each representing the {\revision unknown ``true''} metallicity value of an inner GC. The prior distributions for these are assumed to be normal distributions centred on the measured metallicities, with standard deviations corresponding to uncertainties. 
When evaluating the likelihood function, for all models in Section~\ref{m&s} except Model 1, which does not incorporate metallicity, the {\revision fitted} metallicity parameters were used to partition the GC population, rather than the measured metallicity values. For simplicity, we did not employ a hierarchical model, so there is no tendency for one GC’s true metallicity value to be informed by another GC’s measurement.

\section{Modelling}
\label{m&s}
\subsection{Model 1}
\label{fullmodel}
As discussed above, there are only three unknown parameters in Model 1;
amplitude $A$, orientation $\phi$, and velocity dispersion $\sigma$.
Figure \ref{vm1} shows sample lines drawn from the posterior distribution
to depict the plausible values of the orientation angle $\phi$ of the rotation on the sky.
The orientation angle is estimated to be $37\pm{10}$ degrees
{\revision (PA $=307 \pm 10$ degrees)}.
This is consistent with the direction of Andromeda's stellar disk rotation indicated in \cite{2019Natur.574...69M}. According to Figure \ref{vm1hist}, the posterior distributions of all parameters are approximately Gaussian. Posterior summaries for the parameters are shown in Table \ref{vm1sum}. {\revision The parameter
estimates are the posterior median and the 68\% central
credible interval.}
The posterior distributions indicate that the velocity dispersion
$\sigma$ is almost certainly greater than the rotational amplitude $A$.
This highlights the complexity of the M31 GC system,
as the velocity dispersion acts like a noise term, accounting for deviations
from the $V$ model's assumptions about the rotational pattern. {\revision Quantitatively, we find $A/\sigma = 0.50^{+0.07}_{-0.08}$, suggesting that the GC system is predominantly dominated by dispersion, with a modest but non-negligible contribution from ordered rotation.}
The marginal likelihood estimate for this model is $\ln(Z)= -2141.09$, which, as we shall see, is the lowest marginal likelihood of all the models considered.

\subsection{Model 2}
\label{TwoC}
Model 1, being a simple one-component model, is unlikely to be able to capture all of the complexity of the GC population.
\citet{2019Natur.574...69M} found that two rotating components
are supported for the outer GCs, where each GC was assigned to
one component or the other based on whether it was located on
a substructure or not. Similarly, \citet{2023MNRAS.518.5778L}
split the inner GC population into two components by metallicity,
with the metallicity cutoff being a free parameter,
also finding evidence for two components. Intriguingly, the orientation angles of one of the components
from each study appeared to align.
This alignment was between the Dulais structure (observed as
lower metallicity inner GCs)
and the outer GCs associated with substructures.
This motivated us to create a model that unifies the previous two, assigning GCs to components using a rule based on metallicity (for inner GCs) and substructure status (for outer GCs). Using this model, we can formally test the hypothesis that the
Dulais structure and the \gcsub~outer population share the same rotational
characteristics, and potentially the same origin.
{\revision As in previous studies, we introduce a critical metallicity value, $M_{\rm crit}$, as a free parameter to determine how each GC from the inner population is assigned to one rotational component or the other. 
The prior for $M_{\rm crit}$ is taken to be Uniform$(-2.8, -0.5)$, encompassing the full range of measured inner GC metallicities. {\revis This range is chosen because, after excluding GCs with metallicities $\ge -0.4$, the highest remaining metallicity is $-0.5$, so the range $(-2.8, -0.5)$ covers the whole range of  the metallicity values used in the analysis.} This framework allows us to perform model selection to test different rules for assigning individual GCs to one rotational component or the other.
} In terms of regression models, we are attempting to explain the response
variable (radial velocity) in terms of GC position, along with metallicity
and/or substructure status as the explanatory variables. With the metallicity
cut, the model suddenly changes behaviour as a function of the explanatory
variables. Therefore, this model can be considered to be a change-point
model.

\begin{figure}
\centering
\includegraphics[width=8.8cm]{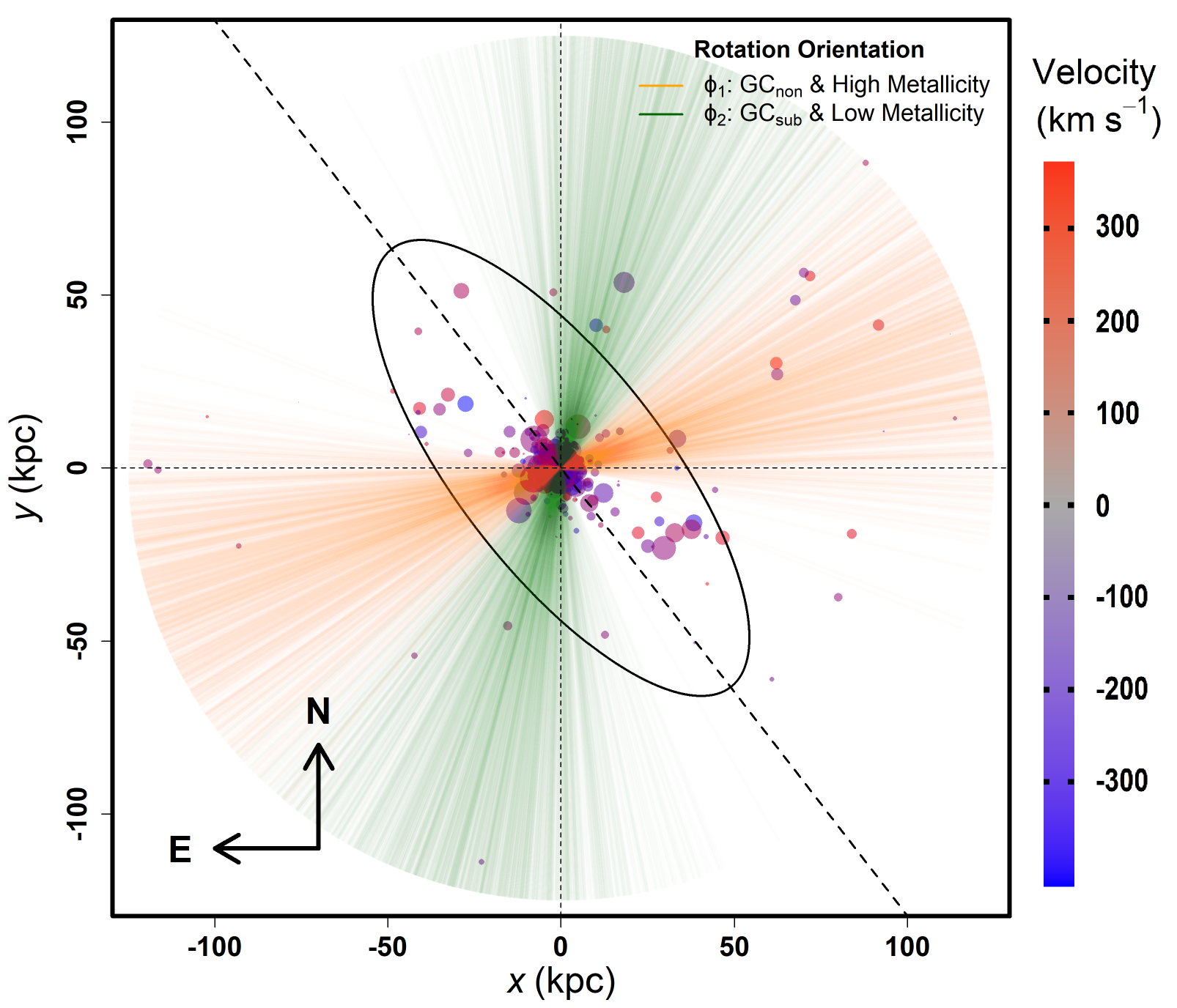}
\caption{As Figure~\ref{vm1}, but for Model 2.1. Here, the \gcsub~and the lower metallicity GCs are grouped together under component 2, with the green lines representing the orientation of the rotational axis drawn from the posterior distribution. The orientation of component 2 agrees with that found in {\revision \citet[][see Figure 2]{2023MNRAS.518.5778L}}, who first suggested the potential association between these two populations.}
\label{vms2}
\end{figure}

In two-component models, the line-of-sight velocity $v$ as a function of position for each component $j \in \{1, 2\}$ is given by
\begin{equation}\label{eqModel}
v_j(x, y) = A_j \sin(\theta - \phi_j).
\end{equation}
Each component has its own rotational amplitude $A_j$ and
orientation angle $\phi_j$, along with its own velocity
dispersion $\sigma_j$ (c.f. Equation~\ref{likelihood}).
The selection of the appropriate component for each GC is
made based on the GC's properties such as its metallicity (for inner GCs)
or its substructure status (for outer GCs).

\begin{table}
	\centering
	\caption{Rules for assigning each GC to one rotating
    component or the other; Models 2.1 and 2.2.}
	\label{scen}
	\begin{tabular}{lcc} 
		\hline
		Model & Conditions & Parameters \\
		\hline
		Model 2.1 & \gcnon~or $\textnormal{Metallicity} > M_{\rm crit}$   & $A_1$, $\phi_1$, and $\sigma_1$ \\ 
           & \gcsub~or $\textnormal{Metallicity} \leq M_{\rm crit}$  & $A_2$, $\phi_2$, and $\sigma_2$   \\ [0.5em]
  
        Model 2.2 & \gcsub~or $\textnormal{Metallicity} > M_{\rm crit}$   & $A_1$, $\phi_1$, and $\sigma_1$ \\ 
           & \gcnon~or $\textnormal{Metallicity} \leq M_{\rm crit}$  & $A_2$, $\phi_2$, and $\sigma_2$   \\ [0.5em]
		\hline
	\end{tabular}
\end{table}

Under a two-component model, we must divide the overall GC population into two
subpopulations. There are two different ways to do this, which 
we call Models 2.1 and 2.2 (Table \ref{scen}).
Model 2.1 is motivated by the observation of \citet{2023MNRAS.518.5778L}
that the lower metallicity inner GCs and the \gcsub~outer GCs appear to share
the same orientation. {\revision Therefore, Model 2.1 assumes a GC belongs to rotational component 1 if it is  \gcnon~or has metallicity above  $M_{\rm crit}$; otherwise, it belongs to rotational component 2 (\gcsub~and metallicity $\leq M_{\rm crit}$ clusters).}

\begin{figure*}
\centering
\includegraphics[width=\linewidth]{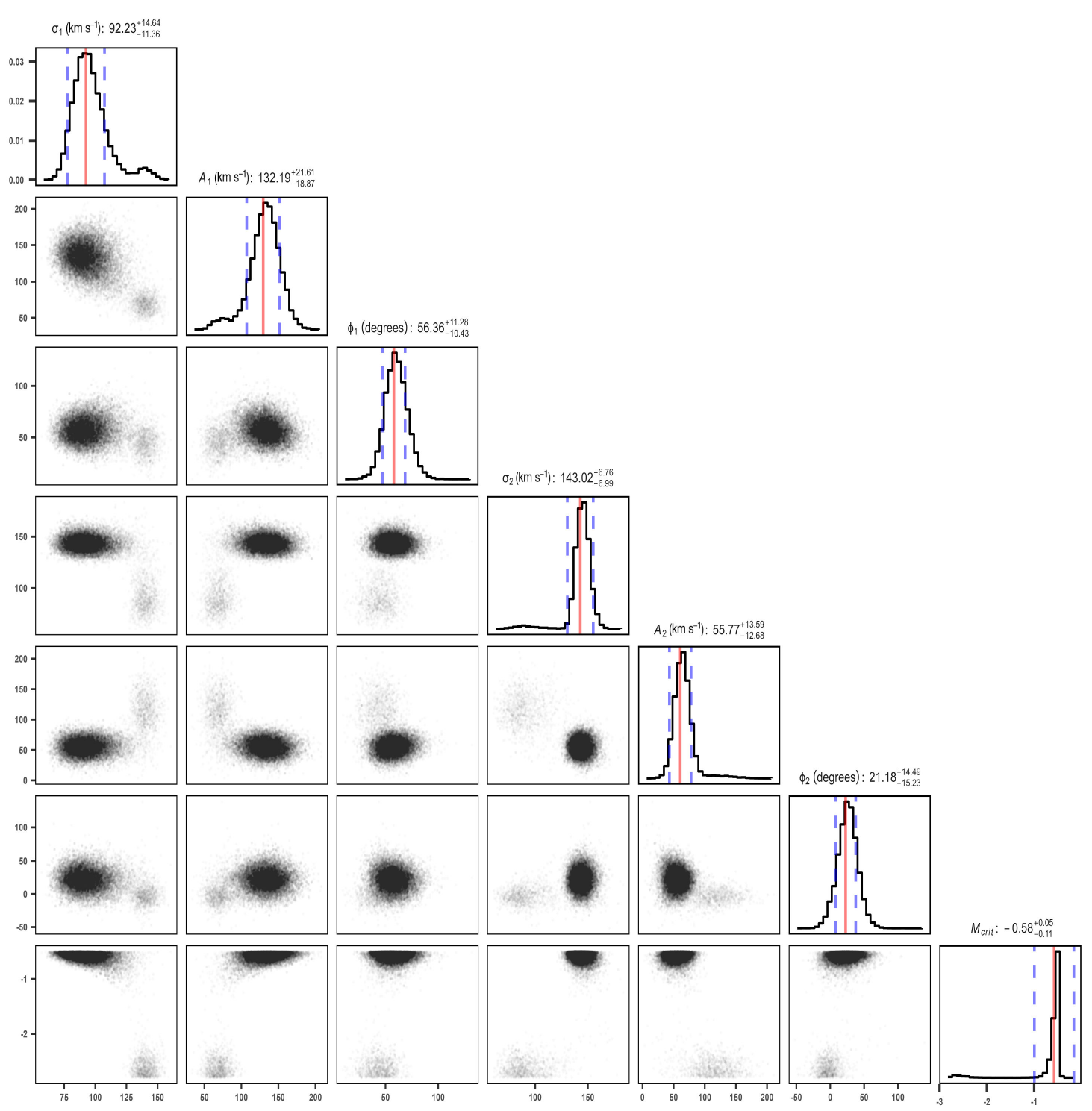}
\caption{Posterior distribution for parameters of Model 2.2. Here, $\sigma_1$, $A_1$ and $\phi_1$ are the parameters for \gcsub~and higher metallicity GCs, while $\sigma_2$, $A_2$ and $\phi_2$ are the parameters for \gcnon~and lower metallicity GCs.}
\label{vmunhist}
\end{figure*}

To determine whether the data indeed support this expectation, we also considered an alternative assumption for the two-component model, which we call Model 2.2. {\revision This model uses swaps the way the GCs are grouped; that is, if a GC is a \gcsub~or higher metallicity GC, it belongs to rotational component 1, otherwise (i.e. if it is a \gcnon~or lower metallicity GC) it belongs to rotational component 2.} These assignment rules are the opposite of those in Model 2.1.
The reasoning behind these ideas is that if galaxies were accreted onto M31 at different epochs, {\revision each of these galaxies would bring its own population of globular clusters. As a result, the GC population in M31} would be separated into various subgroups with different kinematic features.
If there is truly an association between the Dulais structure (inner lower metallicity GCs)
and the outer \gcsub~clusters, we should expect to see Model 2.1 outperform
Model 2.2.

\begin{table}
	\centering
	\caption{Summary table of Model 2.1's parameter estimates.\label{vm2s2sum}}
	\begin{tabular}{lrr} 
		\hline
		Components & Parameters & Estimates \\
		\hline \\[-0.5em]
		  Rotational Component 1 &$\sigma_1$ (\kms )& $137.42^{+6.13}_{-5.82}$ \\[0.5em]
		  & $A_1$ (\kms )& $60.62^{+11.11}_{-11.12}$ \\[0.5em]
            & $\phi_1$ (degrees)& $19.26^{+12.61}_{-12.06}$ \\[0.5em]
            & {\revision PA$_1$ (degrees)}& {\revision $289.26^{+12.61}_{-12.06}$} \\[0.5em]
            & $A_1/\sigma_1$ ratio & $0.44^{+0.08}_{-0.08}$ \\[0.5em]
            \hline \\[-0.5em]
            Rotational Component 2 &$\sigma_2$ (\kms ) & $101.71^{+15.93}_{-16.04}$ \\[0.5em]
            & $A_2$ (\kms ) & $148.03^{+21.57}_{-21.98}$ \\[0.5em]
            & $\phi_2$ (degrees)& $76.69^{+13.69}_{-14.85}$ \\[0.5em]
            & {\revision PA$_2$ (degrees)} & {\revision $346.69^{+13.69}_{-14.85}$} \\[0.5em]
            & $A_2/\sigma_2$ ratio & $1.39^{+0.30}_{-0.27}$ \\[0.5em]
            \hline \\[-0.5em]
            & $M_{\rm crit}$ & $-2.14^{+0.21}_{-0.39}$ \\[0.5em]
		\hline
	\end{tabular}
\end{table}

\begin{figure}
\centering
\includegraphics[width=8.8cm]{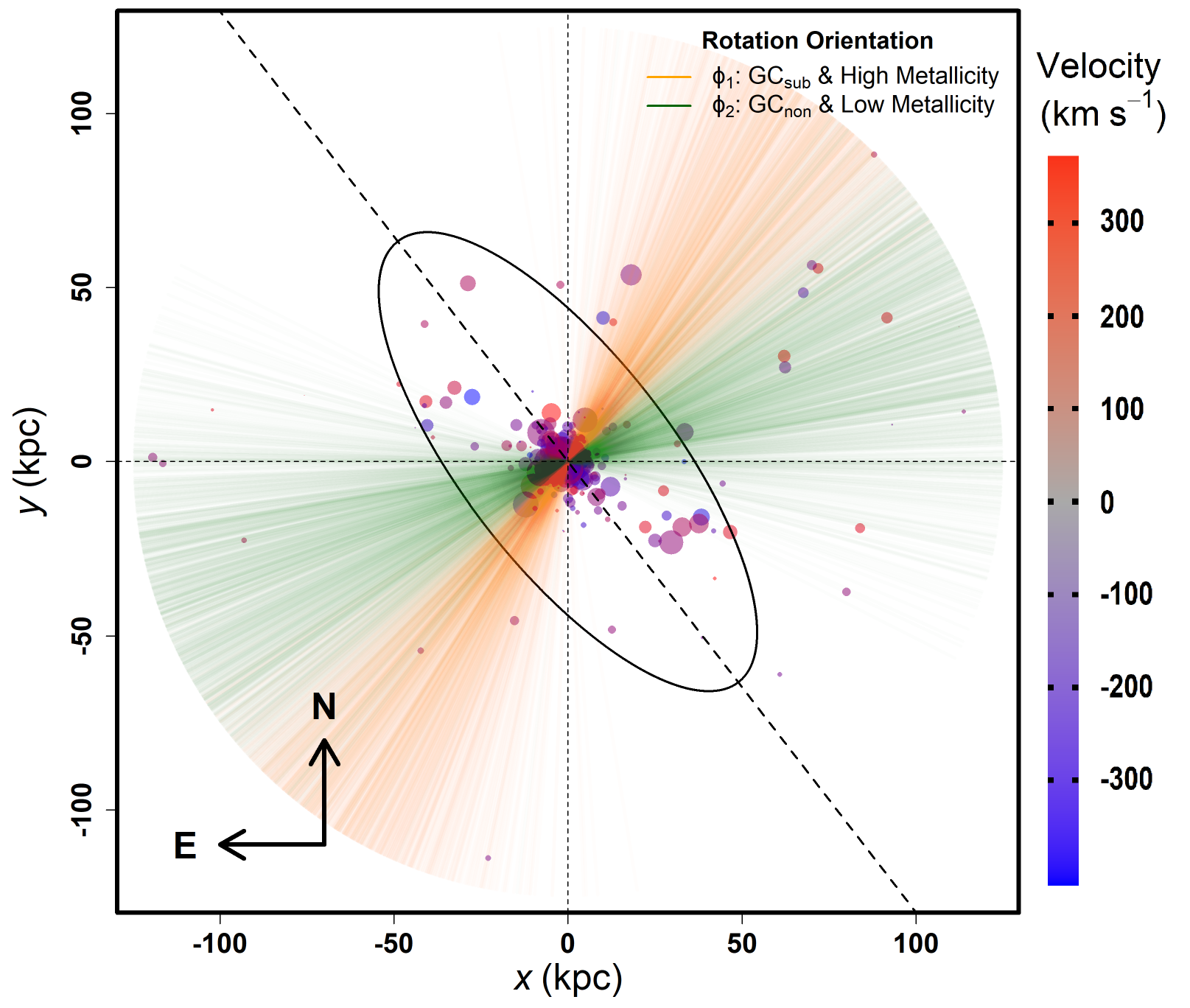}
\caption{As Figure~\ref{vm1}, but for Model 2.2. This assigns \gcsub~and higher metallicity GCs to one component and \gcnon~or lower metallicity GCs to the other. The orientations are significantly different than for Model 2.1; however, {\revision Model 2.2 is less preferred than Model 2.1}. \label{vmun}
} 
\end{figure}

\subsubsection{Model 2.1 Results}
\label{s2}
Under this assumption, \gcnon~ and higher metallicity GCs are grouped together using the parameters $A_1$, $\phi_1$, and $\sigma_1$, while \gcsub~ and lower metallicity clusters share a common rotational component characterized by $A_2$, $\phi_2$, and $\sigma_2$.
Figure \ref{vms2hist} illustrates the posterior distributions of the parameters
for Model 2.1.
Most of the posterior probability for
$M_{\rm crit}$ is below \text{-}1.8 with most of the probability centered around \text{-}2.14, while all other parameters have fairly symmetric posterior distributions. There are {\revision 278} inner GCs of M31 with a metallicity value above \text{-}2.14 and only 9 inner GCs with a metallicity value less than or equal to \text{-}2.14 ---
the proportion of the population with metallicity below $M_{\rm crit}$ is small.
Figure \ref{vms2} demonstrates that the rotational angle,$\phi_1$ (shown with orange lines), for the \gcnon~and high metallicity GCs is $19 \pm 12$ degrees {\revision (PA $= 289 \pm 12$ degrees)}, while the average rotational angle, $\phi_2$ (the green lines), for \gcsub~and lower metallicity GCs is $108 \pm 16$ degrees {\revision (PA $=18 \pm 16$ degrees)}. Lastly, Table \ref{vm2s2sum} shows the posterior summaries for the parameters. The posterior median of $A_2$ (the rotation velocity of lower metallicity and \gcsub~clusters)
is higher than $A_1$ (higher metallicity and \gcnon~clusters).  {\revision The $A_j/\sigma_j$ (or equivalently $V/\sigma$) ratio is calculated for each rotational component that quantifies the balance between ordered rotation ($V$) and random motion ($\sigma$). Systems with higher $A/\sigma$ values are rotation-dominated, whereas those with lower $A/\sigma$ values are dispersion-dominated. Under this assumption, the corresponding kinematic ratios are $A_1/\sigma_1 = 0.44^{+0.08}_{-0.08}$ for the first component and $A_2/\sigma_2 = 1.39^{+0.30}_{-0.27}$ for the second, indicating that the \gcnon~and higher metallicity GC population is predominantly dominated by dispersion, whereas the \gcsub~ and lower metallicity GC population is largely rotation-dominated.}
The marginal likelihood estimate of Model 2.1 is $\ln(Z)= -2139.40$, which
is the highest marginal likelihood of all the models considered.

\begin{table}
	\centering
	\caption{Parameter estimates for Model 2.2.}
	\label{unsum}
	\begin{tabular}{lrr} 
		\hline 
		Components  & Parameter & Estimate  \\
		\hline \\[-0.5em]
		Rotational Component 1 & $\sigma_1$ (\kms)& $92.23^{+14.64}_{-11.36}$  \\[0.5em]
		& $A_1$ (\kms)& $132.19^{+21.61}_{-18.87}$  \\[0.5em]
        & $\phi_1$ (degrees)& $56.36^{+11.28}_{-10.43}$\\[0.5em]
        & {\revision PA$_1$ (degrees)} & {\revision $326.36^{+11.28}_{-10.43}$ }\\[0.5em]
        & $A_1/\sigma_1$ ratio & $1.45^{+0.31}_{-0.34}$ \\[0.5em]
        \hline \\[-0.5em]
        Rotational Component 2 & $\sigma_2$ (\kms)& $143.02^{+6.76}_{-6.99}$ \\[0.5em]
        & $A_2$ (\kms)& $55.77^{+13.59}_{-12.68}$ \\[0.5em]
         & $\phi_2$ (degrees)& $-21.18^{+14.49}_{-15.23}$ \\[0.5em]
         & {\revision PA$_2$ (degrees)} & {\revision $248.82^{+14.49}_{-15.23}$} \\[0.5em]
         & $A_2/\sigma_2$ ratio & $0.39^{+0.10}_{-0.09}$ \\[0.5em]
        \hline \\[-0.5em]
        & $M_{\rm crit}$ & $-0.58^{+0.05}_{-0.11}$  \\[0.5em]
		\hline
	\end{tabular}
\end{table}

\subsubsection{Model 2.2 Results}
\label{unre}
Model 2.2 is the two-component model, but with different assumptions about which subpopulation of inner GCs is associated with which subpopulation of the outer GCs.
For this assumption, \gcsub~and higher metallicity GCs are grouped together with parameters $A_1$, $\phi_1$, and $\sigma_1$. Conversely, \gcnon~and lower metallicity GCs share a common rotational component using $A_2$, $\phi_2$, and $\sigma_2$. {\revision This grouping is the opposite of that suggested by \citet{2023MNRAS.518.5778L}, who found that lower metallicity inner GCs (the Dulais structure) are more plausibly associated with the outer \gcsub~population.}

\begin{figure*}
\centering
\includegraphics[width=\linewidth]{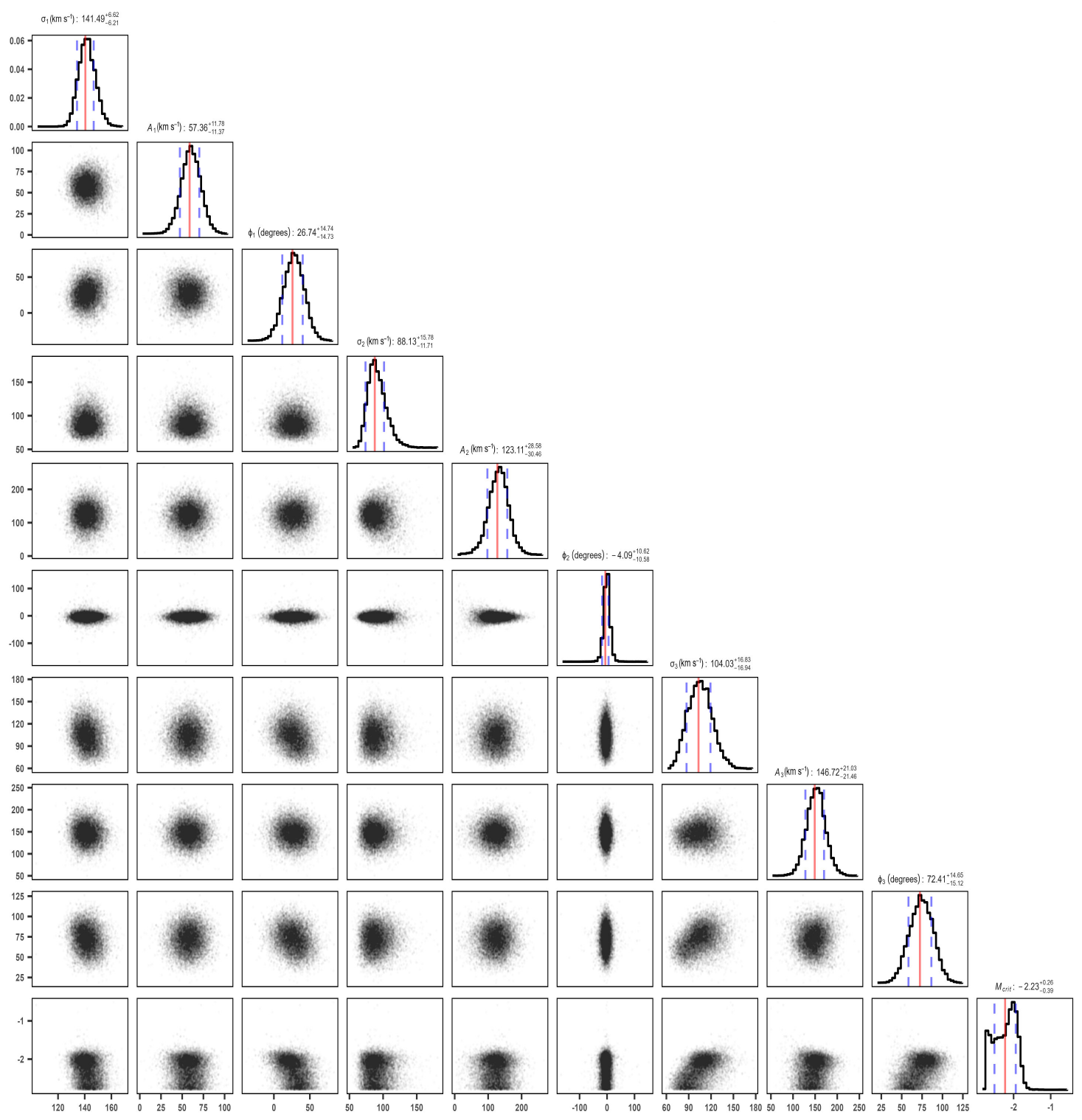}
\caption{Posterior distributions for Model~3. Parameters $\sigma_1$, $A_1$, and $\phi_1$ correspond to the higher-metallicity inner GCs; $\sigma_2$, $A_2$, and $\phi_2$ to the \gcnon~outer clusters; and $\sigma_3$, $A_3$, and $\phi_3$ to the \gcsub~and lower-metallicity GCs.
}
\label{s3hist}
\end{figure*}

Figure \ref{vmunhist} shows the posterior distributions for the parameters
of Model 2.2.
The posterior distribution for $M_{\rm crit}$ is bimodal with most of the probability centred around \text{-}0.58, in this case
the second mode contains only a very small amount of probability. The posterior distribution for $\sigma_1$ and $\sigma_2$ seems to be bimodal, again with the second mode containing only a very small amount of probability.
For all other parameters the posterior distributions are reasonably symmetric.
Figure~\ref{vmun} shows posterior samples for the orientation of rotation 
for the two components. This shows that, under the assumption that the higher metallicity GCs and \gcsub~share a common rotation axis, their orientation angle $\phi_1$ is $56^{+11}_{-10}$ degrees {\revision (PA $=326 ^{+11}_{-10}$ degrees)}. This angle agrees with the orientation of Andromeda's stellar disk from the previous studies \citep{2019Natur.574...69M}. Conversely, $\phi_2$ (the green lines) reveals that, under the assumption that the lower metallicity GCs and \gcnon~share a similar rotation axis, their rotational orientation $\phi_2$ is $-21 \pm 15$ degrees {\revision (PA $=249 \pm 15$ degrees)}. Lastly, Table \ref{unsum} displays the posterior summary statistics for the parameters. {\revision Quantitatively, the kinematic ratios are $A_1/\sigma_1 = 1.45^{+0.31}_{-0.34}$ for the first component and $A_2/\sigma_2 = 0.39^{+0.10}_{-0.09}$ for the second. 
These values indicate that the component associated with the higher metallicity and \gcsub\ clusters is rotation-dominated, whereas the component linked to the lower metallicity and \gcnon\ clusters is primarily dispersion-supported.} Moreover, the marginal likelihood estimate for this scenario is $\ln(Z)= -2141.78$, which is below those of Models 1 and 2.1.

\begin{figure}
\centering
\includegraphics[width=8.8cm]{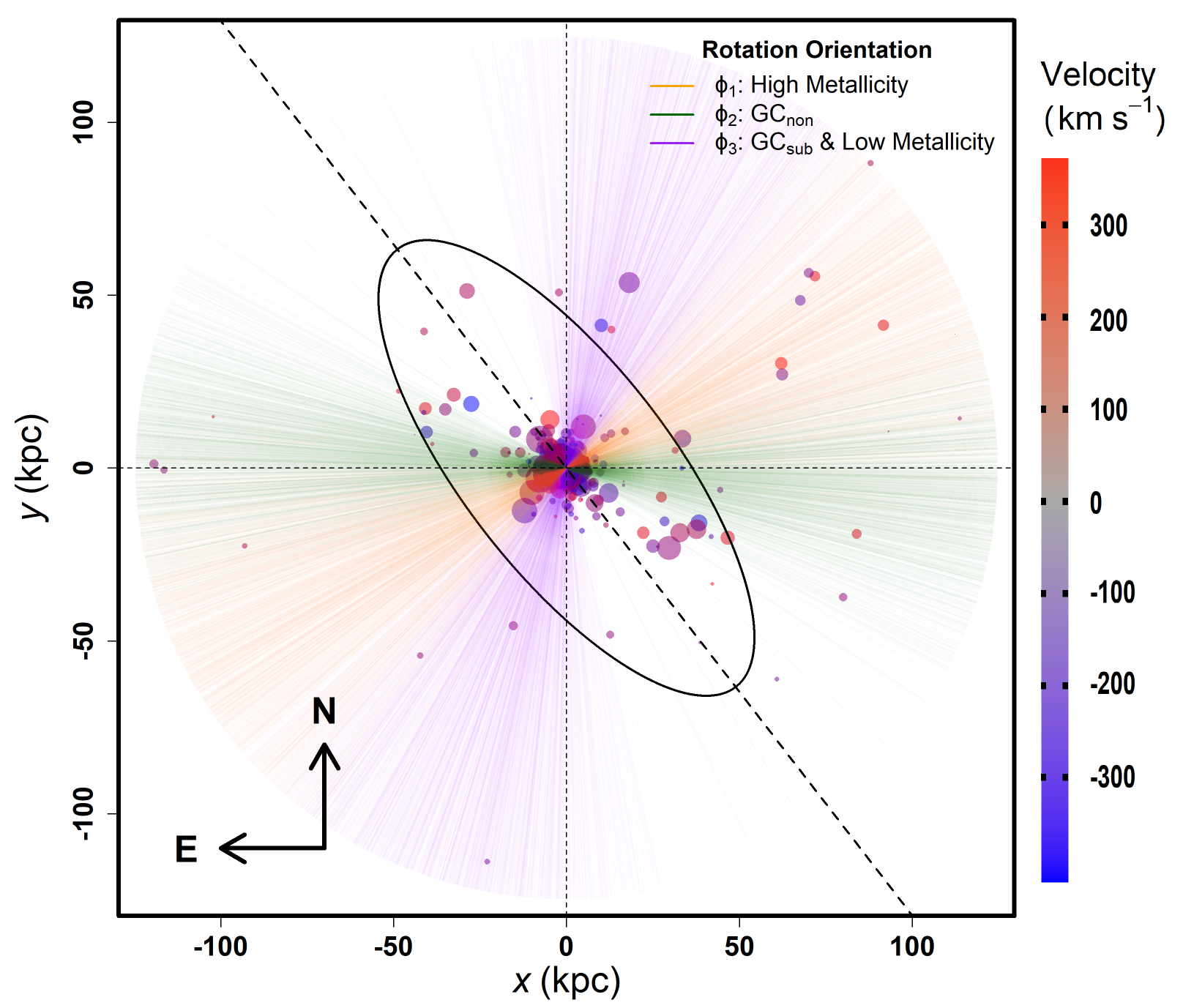}
\caption{As Figure~\ref{vm1} but for Model 3. This assigns higher metallicity inner GCs component 1, \gcnon~to component 2, and \gcsub~or lower metallicity GCs to component 3. The angle of orientation of component 3 agrees with the equivalent component of Model 2.1. The model comparison calculation suggests this model is the second-best model behind Model 2.1.}
\label{s3}
\end{figure}

\subsubsection{Interpretation of the Posterior Distributions for $M_{\rm crit}$}
{The posterior distributions of $M_{\rm crit}$ shown in Figures~\ref{vms2hist} and \ref{vmunhist}, for Models 2.1 and 2.2 respectively, exhibit markedly different features due to the distinct rules each model uses
to assign GCs to rotational components (Table~\ref{scen}).
In Model 2.1, the proposed association between inner and outer GCs is physically motivated as the previous findings from \citet{2023MNRAS.518.5778L} suggest a kinematic connection between
the lower metallicity inner population and the outer \gcsub~clusters.
As a result, the posterior distribution of $M_{\rm crit}$ in Model 2.1 is centred around $-2.14$, effectively separating a small (size $\sim$10) low-metallicity subset of
clusters from the rest.

In contrast, Model 2.2 assigns higher metallicity clusters and \gcsub~clusters to one component, and lower metallicity and \gcnon~clusters to the other. This is less physically motivated,
as there is little previous evidence that \gcnon~outer clusters share a common origin with the lower metallicity inner clusters.
The posterior for $M_{\rm crit}$ under Model 2.2 is bimodal with the most
significant peak close to -0.6. This cutoff includes a much larger fraction of inner GCs in the low metallicity group, effectively reclassifying many GCs
with moderate metallicity into the same kinematic component as the \gcnon~outer population.}

\subsection{Model 3}
\label{ThreeC}
Finally, a three-component model, Model 3, was also considered.
In Model 2.1, we considered the potential association between
the Dulais structure and the \gcsub~outer clusters.
All other GCs were assigned to the other component.
However, it is possible that the connection {\revision between \gcsub~outer clusters and lower-metallicity inner GCs} suggested by
\citet{2023MNRAS.518.5778L} exists, yet
there is no association between the higher metallicity inner clusters and the \gcnon~outer ones.
Therefore, in Model 3, we assume that the population is split into three groups: higher metallicity GCs from the inner population are assigned to the first component. A \gcnon~from the outer clusters is assigned to the second component. Finally, the lower metallicity inner GCs and the \gcsub~outer GCs are assigned to the third component (Table~\ref{model3}). {\revision This grouping is physically motivated, as previous studies \citep{2023MNRAS.518.5778L} suggest a connection between \gcsub~outer GCs and low-metallicity populations.}
This final component is the same as in Model 2.1 and is the one highlighted throughout this paper.

\begin{table}
	\centering
	\caption{Rules for assigning GCs to subgroups for Model 3.}
	\label{model3}
	\begin{tabular}{lc} 
		\hline
		Conditions & Rotating Components \\
		\hline
         $\textnormal{Metallicity} > M_{\rm crit}$  & $A_1$, $\phi_1$, and $\sigma_1$ \\
        \gcnon  & $A_2$, $\phi_2$, and $\sigma_2$\\ 
        \gcsub~or $\textnormal{Metallicity} \leq M_{\rm crit}$ & $A_3$, $\phi_3$, and $\sigma_3$\\ [0.5em]
		
	\end{tabular}
\end{table}

For the three-component model,
the line-of-sight velocity $v$ as a function of position for each component $j \in \{1, 2, 3\}$ {\revision has} a form given by Equation~\ref{eqModel}.
Each of the three components has its own rotational amplitude $A_j$, orientation angle $\phi_j$, and velocity dispersion $\sigma_j$. As with the previous models, the velocity dispersion contributes to the variance term in the likelihood function. 

\begin{table}
	\centering
	\caption{Parameter estimates for Model 3.}
	\label{s3sum}
	\begin{tabular}{lrr} 
		\hline
		Components & Parameters & Estimates\\
		\hline
		Rotational Component 1 & $\sigma_1$ (\kms)& $141.49^{+6.62}_{-6.21}$ \\[0.5em]
		& $A_1$ (\kms)& $57.36^{+11.78}_{-11.37}$ \\[0.5em]
        & $\phi_1$ (degrees)& $26.74^{+14.74}_{-14.73}$ \\[0.5em]
        & {\revision PA$_1$ (degrees)}& {\revision $296.74^{+14.74}_{-14.73}$} \\[0.5em]
        & $A_1/\sigma_1$ ratio & $0.41^{+0.08}_{-0.09}$ \\[0.5em]
        \hline \\[-0.5em]
        Rotational Component 2 & $\sigma_2$ (\kms)& $88.13 ^{+15.78}_{-11.71}$\\[0.5em]
       &  $A_2$ (\kms)& $123.11 ^{+28.58}_{-30.46}$ \\[0.5em]
        & $\phi_2$ (degrees)& $-4.09^{+10.62}_{-10.58}$ \\[0.5em]
        & {\revision PA$_2$ (degrees)}& {\revision $265.91^{+10.62}_{-10.58}$} \\[0.5em]
         & $A_2/\sigma_2$ ratio & $1.40^{+0.40}_{-0.41}$ \\[0.5em]
        \hline \\[-0.5em]
        Rotational Component 3 & $\sigma_3$ (\kms)& $104.03^{+16.83}_{-16.94}$ \\[0.5em]
       & $A_3$ (\kms)& $146.72^{+21.03}_{-21.46}$ \\[0.5em]
       & $\phi_3$ (degrees)& $72^{+14.65}_{-15.12}$ \\[0.5em]
       & {\revision PA$_3$ (degrees)}& {\revision $342^{+14.65}_{-15.12}$} \\[0.5em]
       & $A_3/\sigma_3$ ratio & $1.42^{+0.31}_{-0.28}$ \\[0.5em]
       \hline \\[-0.5em]
        & $M_{\rm crit}$ & $-2.23 ^{+0.26}_{-0.39}$ \\[0.5em]
		\hline
	\end{tabular}
\end{table}

Figure \ref{s3hist} illustrates that the posterior samples of $M_{\rm crit}$ are below {\revision $[\textnormal{Fe}/\textnormal{H}] = -1.7$}, with most of the possibilities centered around {\revision $[\textnormal{Fe}/\textnormal{H}] = -2.23$} (in agreement with Model 2.1), while all other parameters have approximately normal
posterior distributions. Figure~\ref{s3} illustrates that the average rotating angle $\phi_1$ (the orange lines) for the higher metallicity inner GCs, is
$27 \pm 15$ degrees {\revision (PA $=297 \pm 15$ degrees)}.
The orientation angle $\phi_2$ (the green lines), for the \gcnon~is $-4\pm 11$ degrees {\revision (PA $=266 \pm 11$ degrees)}, and $\phi_3$ (the purple lines) for the \gcsub~and lower-metallicity GCs is $72 \pm 15$ degrees {\revision (PA $=342 \pm 15$ degrees)}. This latter result is consistent with the conclusion of Model 2.1. {\revision Interestingly, the higher metallicity inner GCs in Model 3 rotate in the same direction as M31's stellar disk, consistent with previous findings \citep{2019Natur.574...69M}.} {\revision Quantitatively, the kinematic ratios are $A_1/\sigma_1 = 0.41^{+0.08}_{-0.09}$ for the first component, $A_2/\sigma_2 = 1.40^{+0.40}_{-0.41}$ for the second, and $A_3/\sigma_3 = 1.42^{+0.31}_{-0.28}$ for the third. These results indicate that the higher metallicity inner GCs (component~1) are primarily dispersion supported, whereas both the \gcnon\ (component~2) and the \gcsub\ together with lower metallicity GCs (component~3) exhibit rotation dominated kinematics.} The posterior summaries for the parameters are shown in Table \ref{s3sum}.
The marginal likelihood estimate for Model 3 is $\ln(Z)= -2141.03$,
which is the second highest of the four models considered in this paper,
behind Model 2.1.

\subsection{Bayesian evidence comparison}
\label{bec}
The marginal likelihoods (evidence) for the models are given in Table~\ref{3sc},
along with the Bayes Factors relative to the best model (Model 2.1).
The following equation provides an example of determining the Bayes Factor, in this case for Model 2.1 versus Model 2.2:
\begin{equation}\label{exeq}
    \begin{split}
     \textnormal{Bayes Factor} &= \\
     \frac{Z_{2.1}}{Z_{2.2}} & = \exp\left(\log\mathcal{P}(\mathcal{D}\,|\,\mathcal{M}_{2.1}) - \log\mathcal{P}(\mathcal{D}\,|\,\mathcal{M}_{2.2})\right)\\ 
     & = \exp(-2139.40 - (-2141.78)) \\
     & \approx 11.
    \end{split}
\end{equation}

Following \cite{Kass}, a Bayes Factor ranging from 3.2 to 10 indicates substantial evidence in favour of one model over another, and strong evidence is indicated by a Bayes factor between 10 and 100. Based on this, Bayes Factors of approximately 5, 11, and 6 for Model 2.1 versus the alternatives indicate that the data provide substantial to strong support for Model 2.1. The Bayes Factor further indicates that the alternative assumption for the two-component model (Model 2.2), indeed, is not a plausible model.
Table~\ref{3sc} shows the evidence ratios relative to Model 2.1, this time with the
ratio inverted so that the Bayes Factors are less than one.
Moreover, if the prior probabilities of each of the four models
are all equal to 1/4, the posterior probabilities become
$(0.12, 0.68, 0.06, 0.13)$ respectively.
There is significant plausibility for Model 2.1 and some plausibility
for Model 3, whose implications are similar to Model 2.1, in that they
both assume the association suggested by \citet{2023MNRAS.518.5778L}.
Model 1, while disfavoured by the data, is not completely ruled out.

\begin{table}
	\centering
	\caption{The marginal likelihood estimates for each of the four models
    considered in this paper.}
	\label{3sc}
	\begin{tabular}{lcc} 
		\hline
		Model & $\ln{Z}$ & $Z/Z_{\rm max}$\\
		\hline
            Model 1  & $-2141.09$ & $0.18$\\ 
            Model 2.1 & $-2139.40$ & $1.00$ \\ 
            Model 2.2  & $-2141.78$ & $0.09$\\ 
            Model 3  & $-2141.03$ & $0.20$\\ 
		\hline
	\end{tabular}
\end{table}

\section{Discussion and Conclusions}
\label{Dis}
Here, we have explored various hypotheses about the kinematic structure
of M31's GC population. Motivated primarily by previous separate studies
of the inner \citep{2023MNRAS.518.5778L} and outer \citep{2019Natur.574...69M} GC populations, we have combined the two populations and performed a unified analysis. Since the evidence supports two distinct subpopulations for both the inner
and outer GCs when analysed separately, we investigated the possibility that
one of the inner subpopulations is associated with one of the outer subpopulations, {\revis which could occur} as a result of being accreted onto M31 as part of one accretion event. {\revision It should be noted that, as mentioned in Section~\ref{Data}, photometric metallicity estimates (and a few spectroscopic ones) are now available for the outer GCs \citep{2025MNRAS.542L..60M}. However, due to the heterogeneous origin and incomplete coverage of the outer GC sample, these metallicities were not included in the modelling process.}

Model 1 (Subsection~\ref{fullmodel}) provides a baseline, showing that the overall rotation axis of the GC population aligns with Andromeda’s stellar disk. However, this model is not well supported in the final model comparison. In Models 2.1 and 2.2, we divide the GCs into two components, by metallicity for inner GCs and by substructure association for outer GCs, to test different assumptions about their shared origins. These models explore whether inner and outer GCs could have been deposited by the same progenitor galaxies during separate accretion events. Following \citet{2023MNRAS.518.5778L}, Model 2.1 assumes that lower-metallicity inner GCs and substructure-associated outer GCs form a single population. This hypothesis is strongly supported by the data and predicts a {\revision rotation axis that is nearly north-south on the sky}, consistent with previous findings (see Figure~\ref{vms2}). In contrast, Model 2.2, which assumes the opposite association, is disfavoured. Table~\ref{vm2s2sum} {\revision (for Model 2.1)}
shows that the posterior for amplitude $A_1 \sim 50-70$ \kms, is significantly lower than for $A_2$ with a velocity of $\sim125-170$ \kms, implying that the lower-metallicity or substructure GCs rotate about 2.4 times faster than their higher-metallicity or non-substructure counterparts.
{\revision  It appears that the lower-metallicity or substructure GCs rotate faster but also have a lower velocity dispersion. Considering the velocity dispersion (see Table~\ref{vm2s2sum}), the corresponding kinematic ratios are $A_1/\sigma_1 = 0.44^{+0.08}_{-0.08}$ for the higher-metallicity or non-substructure GCs and $A_2/\sigma_2 = 1.39^{+0.30}_{-0.27}$ for the lower-metallicity or substructure GCs. This indicates that the higher-metallicity or non-substructure GCs are predominantly dispersion supported, whereas the lower-metallicity or substructure GCs are largely rotation dominated. {\revis Note that this comparison excludes GCs not studied
in this paper --- those with metallicities
greater than or equal to $-0.4$ --- which exhibit coherent rotation aligned with the stellar disk.}
The stronger coherent rotation of the lower-metallicity {\revis and } substructure GCs may reflect the accretion history of M31, where these clusters, ({\revis possibly due to having been accreted more recently}), retain angular momentum from their progenitor systems, while higher-metallicity clusters have been dynamically heated and partially isotropized.}


Finally, a three-component model (Model 3) was also explored to further examine the relationships between the different GC subgroups. 
This model, which seeks to identify further substructures within the GC population, whilst not as strongly supported as Model 2.1, is the second most plausible model considered.
Additionally, the result from Model 3 demonstrates that the higher metallicity inner clusters rotate in the same direction as Andromeda's stellar disk, consistent with previous studies \citep{2019Natur.574...69M}. The physical implications of Model 3 and Model 2.1, the two best-supported models, are similar, and both support the proposed association of \citet{2023MNRAS.518.5778L}.
{\revision The existence of metallicity estimates for a subset of the outer
GCs raises the question of whether there is a metallicity difference
between the \gcsub~and \gcnon~outer populations. A naive extrapolation of our
results might suggest that the \gcsub~population should typically have lower metallicity. \citet{2025MNRAS.542L..60M} stated that they did not detect any substantial metallicity
difference between these two populations. However, their Figure 3
suggests that there might indeed be slightly more low-metallicity GCs in the
substructure group. Testing this possibility formally would require
incorporating these additional
measurements into the analysis would require significant and complex changes to the
models.}


As a closing remark, these results highlight the complexity of M31’s accretion history, revealing that its GC system comprises dynamically and chemically distinct subpopulations likely deposited during multiple accretion events. To further unravel this history, more metallicity measurements in the outer halo are essential, particularly to clarify the nature and extent of structures like the Dulais. Additionally, accurate distance estimates are needed to reconstruct the full three-dimensional spatial distribution of clusters, as the Dulais Structure is likely to lie along the line of sight to the central regions of M31. Hence, future photometry and spectroscopy of the GC populations will be essential in unravelling the accretion history of Andromeda.

\section*{Acknowledgments}
We are grateful to Annette M. N. Ferguson (Edinburgh) and Daniel Zucker (Macquarie) for insight on earlier versions of this work and comments that improved the presentation of the results. We also thank the two anonymous referees
and the editor
who provided useful feedback.

\bibliographystyle{mnras}
\bibliography{references} 



\appendix

\section{Appendix A: Other Kinematic Models}
\label{Models}
In this study, we analysed the data under the assumption of the
$V$ model, a specific functional form for the radial velocity
as a function of position on the sky.
Here, we enlarge the set of possible functions by considering
an $S$ (solid body) and $F$ (asymptotically flat) model as well.
The functional form of each rotational model
is displayed in Table \ref{models}, and below is the description of each model:

\begin{itemize}
    \item 
    Model $V$ was presented by \citet{Cote_2001} as well as \citet{2014MNRAS.442.2929V}. This model considers a line-of-sight velocity profile that varies smoothly as a function of angle. This model
    was assumed throughout the main part of the paper.
    \item
    Model $S$ denotes a solid-body rotation of the GC population, in which the velocity grows linearly with distance from the rotation axis.
    \item
    Model $F$ represents an asymptotically flat rotation of the GC
    population. It assumes that line-of-sight velocity changes from one side of the rotation
    axis to the other, with a smooth transition in between (which may be a slow or a fast transition). The length scale of the transition is determined by an extra
    parameter $L$.
    For the prior on $L$, we used $\log_{10}(L/\textnormal{kpc}) \sim \textnormal{Uniform}(-1, 2)$ for each component.
\end{itemize}

\begin{table}
	\centering
	\caption{Line-of-sight velocity of models $V$, $S$, and $F$.}
	\label{models}
	\begin{tabular}{lr} 
            \hline
             Model & Functional form \\
            \hline
            $V$  & $v_{r}(x_i, y_i) = A \sin(\theta_i - \phi)$ \\
            $S$  & $v_{r}(x_i, y_i) = A (x_i\sin(\phi) - y_i\cos(\phi))$ \\
            $F$  & $v_{r}(x_i, y_i) = A \tanh((x_i\sin(\phi) - y_i\cos(\phi))/L)$ \\
            \hline
	\end{tabular}
\end{table}


The marginal likelihood estimates for Models 1, 2.1, 2.2, and 3 for the three different kinds of velocity profiles are shown in Table~\ref{M31fullmc}.
This shows a slight preference for model $F$ over model $V$ except for Model 3,
which favours model $V$. The results also indicate
that kinematic models $V$ and $F$ are vastly preferred over
$S$.

However, for pragmatic reasons, we decided to prioritize Model $V$ 
in the main part of the paper, similar to
\citet{2019Natur.574...69M} but different from \citet{2023MNRAS.518.5778L},
who presented model $F$. One reason for this is that, overall, Model $V$
has a greater precedent in the
literature. Another is that the posterior distributions are significantly
more complex under the $F$ model, due to the sudden transition from one
side of M31 to the other when the $L$ parameter is small.
This complicates the interpretation and the summaries of the posterior
distributions significantly.
Importantly, none of the substantial conclusions of the paper
(the evidence for Models 2.1 and 3 being the highest, and the inferred
orientation angles of the rotational components)
are affected by this choice.

\begin{table*}
	\centering
	\caption{Marginal likelihood $\ln{Z}$ estimates for the three  kinematic models. }
	\label{M31fullmc}
	\begin{tabular}{lcccc} 
		\hline
		Model & Model 1 &  Model 2.1 &  Model 2.2 &  Model 3\\
		\hline
            $V$  & $-2140.09$ & $-2139.40$ & $-2141.78$ & $-2141.03$\\ 
            $S$  & $-2154.63$ & $-2161.22$ & $-2161.05$ & $-2165.38$\\ 
            $F$  & $-2139.83$ & $-2138.23$ & $-2141.23$ & $-2141.53$\\ 
		\hline
	\end{tabular}
\end{table*}

\end{document}